# Electromagnetic and vacuum tests of the PTAK-RFQ module 0

A. Kılıçgedik[1], A. Adıgüzel[2,8], A. Çağlar[3], E. Çelebi[4], Ş. Esen[2], M. Kaya[1],
Ü. Kaya[5], V.E. Özcan[4,8], G. Türemen[5], N.G. Ünel[6] and F. Yaman[7,9]

[1]Marmara University, Department of Physics, İstanbul
[2]İstanbul University, Department of Physics, İstanbul
[3]Yıldız Technical University, Electronics and Communication Engineering, İstanbul
[4]Boğaziçi University, Department of Physics, İstanbul
[5]Turkish Energy, Nuclear and Mineral Research Agency- Nuclear Energy Research Institute (TENMAK-NUKEN), Ankara
[6]University of California Irvine, Physics Department, Irvine
[7]İzmir Institute of Technology (IZTECH), İzmir
[8]Boğaziçi University, Feza Gürsey Center of Physics and Mathematics, İstanbul
[9]Science and Technology Facilities Council(STFC/DL/ASTeC) Daresbury Laboratory Accelerator Science and Technology Center, Daresbury

August 03, 2023

**Abstract**

A new Radio-frequency quadrupole (RFQ), which operates at 800 MHz high frequency and will enable to accelerate of the proton beam efficiently was designed at KAHVELab (Kandilli Detector, Accelerator and Instrumentation Laboratory) at Boğaziçi University in İstanbul, Turkey. The so-called PTAK-RFQ, which consists of two modules with a total length of less than one meter will accelerate protons to 2 MeV at the Proton Testbeam at the Kandilli campus, known as the PTAK project. The prototype of the first module of the 800 MHz PTAK-RFQ (called the PTAK-RFQ module 0), which captures and bunches the proton beam injected from the ion source was fabricated by a local manufacturer from ordinary copper material. The PTAK-RFQ module 0 was subjected to various tests to ensure that its mechanics, pressure, field distribution, and frequency are operationally adjusted. The facilitating solutions emerging from the detailed testing of the PTAK-RFQ module 0 will ultimately guide all mechanical, vacuum, rf testing, final design, and manufacturing processes of the final PTAK-RFQ. The PTAK-RFQ module 0 was first subjected to vacuum tests and then to detailed vacuum leak tests. Subsequently, low-power rf measurements were performed for tuning of field and frequency. The tuning algorithm developed by CERN was optimized for 16 tuners and 6 test field points to be adjusted to the PTAK-RFQ module 0 to the desired field distribution. The tuning algorithm is based on a response matrix, whose inputs are created by bead-pull measurements of individual tuner movements. The tuning algorithm gives some predictions for corrective tuner movements to achieve desired field distribution. In the framework of all these RF tuning processes, the field distribution was tuned through the tuning algorithm and then the frequency was tuned manually.

**INTRODUCTION**

The first phase of the PTAK project, including the low-energy transmission line, has recently been commissioned at KAHVELab in Turkey [1-4]. In line with the PTAK project's ultimate goals, the PTAK-RFQ was designed [5] to be accelerated protons to 2 MeV to be used in Proton Induced X-Ray Emission (PIXE) experiments, a non-destructive elemental analysis technique [6]. Layouts of the ion source, the low-energy beam transport line, and the 800 MHz four-vane RFQ at KAHVELab can be seen in Fig. 1 [2, 7].

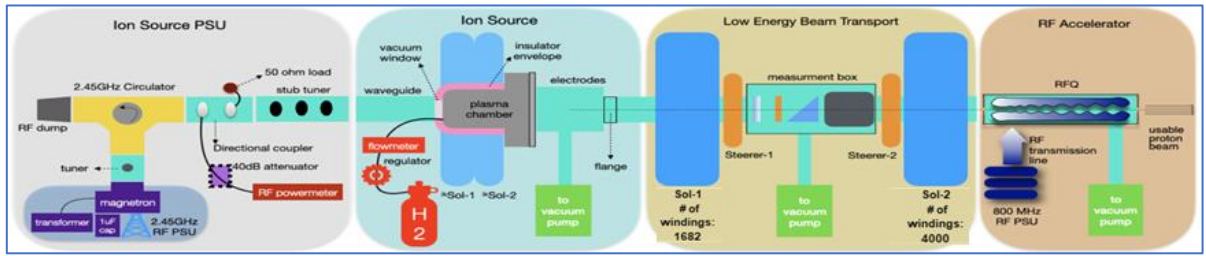

Figure 1: Layout of the Proton Test Beam Line at KAHVELab.

The 800 MHz four-vane PTAK-RFQ is the key accelerator component which consists of two modules with a total length of less than one meter that will accelerate protons to 2 MeV. The PTAK-RFQ is assembled using bolts and nuts without the need for brazing. In addition, PTAK-RFQ's design includes 3-D O-rings to prevent vacuum leaks and finger-type RF shields to resist RF leaks [5].

The PTAK-RFQ and similar high-frequency (HF) RFQs, recently developed by CERN [8 -16], are summarized in terms of their key parameters in Table 1. The PTAK-RFQ will be operated at a higher operating frequency than current similar RFQs and it will come into operation soon as the world's smallest RFQ at KAHVELab in Turkey.

Perfect operation of an RFQ is ensured by minimizing its manufacturing flaws and misalignments [17]. The prototype of the first module of the 800 MHz PTAK-RFQ (called the PTAK-RFQ module 0) was successfully produced from ordinary copper material by a local producer to minimize the problems that may occur in the production and testing processes of the PTAK-RFQ and to investigate its domestic manufacturability. The PTAK-RFQ module 0 was subjected to a series of mechanical, vacuum leak, and low-level RF tests to adjust to desired operational settings in the second quarter of 2022.

The field tuning of any RFQ has remarkable importance in reaching the desired field distribution. The field of the PTAK-RFQ module 0 has been tuned by optimizing some inputs of the equation based on the tuning algorithm developed by CERN for the HF-RFQ [12, 13]. The tuning algorithm is derived from a response matrix [17]. Effects of individual tuner movements on the field are measured by bead-pull measurements and then entries of the response matrix are created by these measurements. Possible tuner corrections are then calculated by inverting the response matrix to obtain the desired field distribution.

Table 1: A Quick Comparison of the PTAK-RFQ at KAHVELab with similar HF RFQs [12,16].

| Parameter | Symbol | HF | PIXE | PTAK |
|---|---|---|---|---|
| Input E (keV) | $W_{in}$ | 40 | 20 | 20 |
| Output E (MeV) | $W_{out}$ | 5 | 2 | 2 |
| RF (MHz) | $f_0$ | 750 | 750 | 800 |
| # of modules | - | 4 | 2 | 2 |
| RFQ length (mm) | - | 1964 | 1072.938 | 980 |
| Quality Factor | $Q_0$ | 6440 | 5995 | 7036 |
| RF Power Loss (kW) | $P_0$ | 350 | 64.5 | 48.5 |

This study reports detailed results of the initial vacuum tests and low-power RF measurements performed to set the desired field distribution and the operating frequency of the PTAK-RFQ module 0.

**THE PTAK-RFQ MODULE 0**

**1. The Initial Vacuum Tests**

The PTAK-RFQ module 0 has been subjected to some vacuum sealing tests to observe whether it has appropriate vacuum holding capabilities. The special 3-D vacuum seal, which is shown in Fig. 2(a), was designed [5] to prevent vacuum leaks. It is formed by two circular O-rings and four flat pieces from Viton material (see Fig. 2(b)), which are affixed with an adhesive suitable for vacuum conditions. The pressure level of the PTAK-RFQ module 0 was measured by operating two turbomolecular pumps in the setup shown in Fig. 2(c).

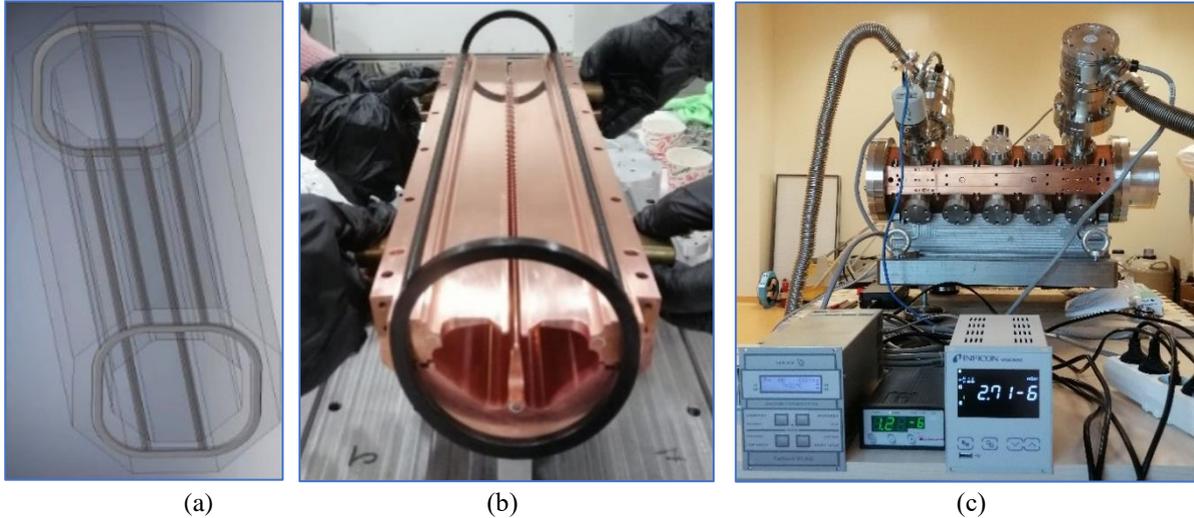

(a)            (b)            (c)

Figure 2: (a) Solid model design of the special 3-D vacuum seal, (b) view of the special 3-D vacuum seal, and (c) photograph of the setup for the initial vacuum tests of the PTAK-RFQ module 0.

During the initial vacuum tests, the turbomolecular pumps were initially operated for 3 hours, after this period, the pressure reached approximately $1.2 \times 10^{-6}$ mbar (Fig. 3). Additionally, the helium leak test was performed with the help of a helium leak detector. No vacuum leak was detected in the PTAK-RFQ module 0 during the detailed helium leak testing process.

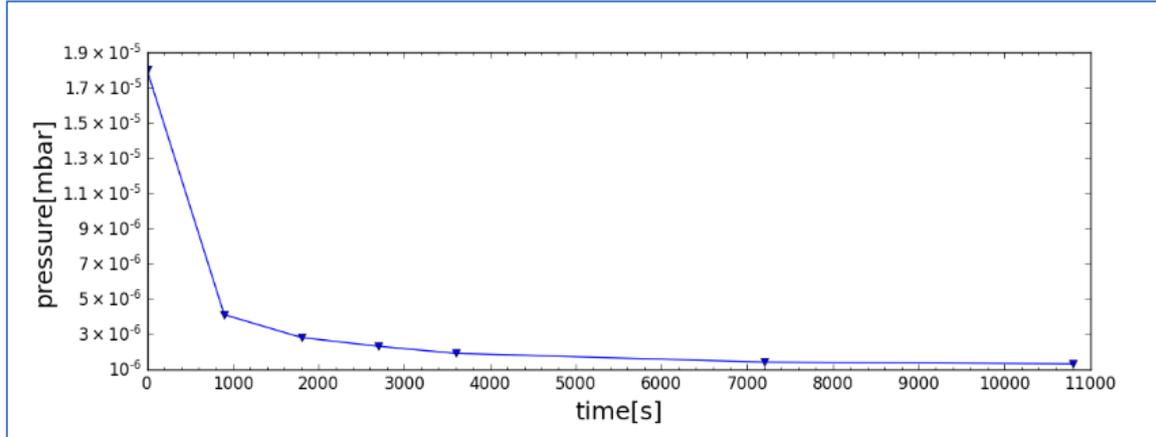

Figure 3: Change of vacuum levels over time for the test module after 100% operation of the two turbomolecular pumps.

Figure 3 shows that the measured pressure is sufficient for the PTAK-RFQ module 0, whereas the vacuum level required for the final PTAK-RFQ to operate has not yet been reached. It was observed that the fast-corroding surfaces of the PTAK-RFQ module 0, which is fabricated of ordinary copper material, need a more detailed cleaning to achieve a better vacuum level. A better pressure level can be expected after thorough cleaning of the PTAK-RFQ module 0. These cleaning steps are: first a deep cleaning with a new organic copper surface cleaning agent (developed by a local manufacturer), then polishing with diamond polish and finally cleaning with the isopropyl alcohol.

Although the cleaning prescription for the PTAK-RFQ module 0 was fulfilled, it was not subjected to a new vacuum test as the current measured pressure was satisfactory.

**2. RF Tests for the Field and the Frequency Tuning of the PTAK-RFQ Module 0**
*2.1. Bead Pull Measurements and Instruments*

The standard bead-pull method was used to measure the longitudinal squared magnetic-field distribution near the outer wall of the PTAK-RFQ module 0 by using a metallic bead. The metallic bead attached to the fishing line was inserted into the PTAK-RFQ module 0 via a stepper motor in small steps. Slater's perturbation theorem [18, 19] states that a small bead volume is subtracted from the total cavity volume at each entrainment step, and so the perturbation in any bead position causes a shift in the resonant frequency of the cavity proportional to the square of the local field [17].

A series of measurements were performed to find the mechanical adjustments for each of the tuners to provide the desired magnetic field distribution in the quadrupole operating mode. The hardware of the bead-pull system consists of the VNA (Vector network analyzer), the metallic bead, the fishing line, N-type cables, two pick-up antennas, pulleys, and the stepper motor. The same VNA, N-type cables (50Ω), and calibration kit (50Ω) were used to ensure equal conditions for results to be comparable throughout all bead-pull measurements. Two homemade pick-up antennas (50Ω), which were manufactured using 2.5mm thick copper wire were placed in the middle of two opposite surfaces of the PTAK-RFQ module 0, which has an octagonal structure.

Bead-pull measurements for the PTAK-RFQ module 0 were performed using the phase $\phi$ of the transmission coefficient $S_{21}$ measured through two pick-up antennas. The experimental setup used for bead-pull measurements of the PTAK-RFQ module 0 is shown in Fig.4. A special tuner kit has been developed and manufactured that allows the tuners to be inserted or retracted into the PTAK-RFQ module 0 up to a precision of 10 micrometers. The solid model design of the special tuner kit is shown in Fig. 5(a) and the manufactured tuner kits are shown in Fig 5(b). They have a locked cover system; thus, the tuners' self-positioning in the PTAK-RFQ module 0 is prevented during each bead-pull measurement. Tuners consisting of copper slugs were arranged in a clockwise spiral, starting from where the protons accelerate from low energy to high energy, as shown in Fig. 6.

Since each vane is designed to be mounted without brazing, i.e., using bolts and nuts, it must be provided with strong protection in terms of RF leaks. Thus, it was decided to use the Copper C-shaped finger-type RF shields in channels at the junction points of each of the conductive vanes as shown in Figs. 7 (a), (b), and the dimensions of the RF shield are shown in Fig. 7 (c).

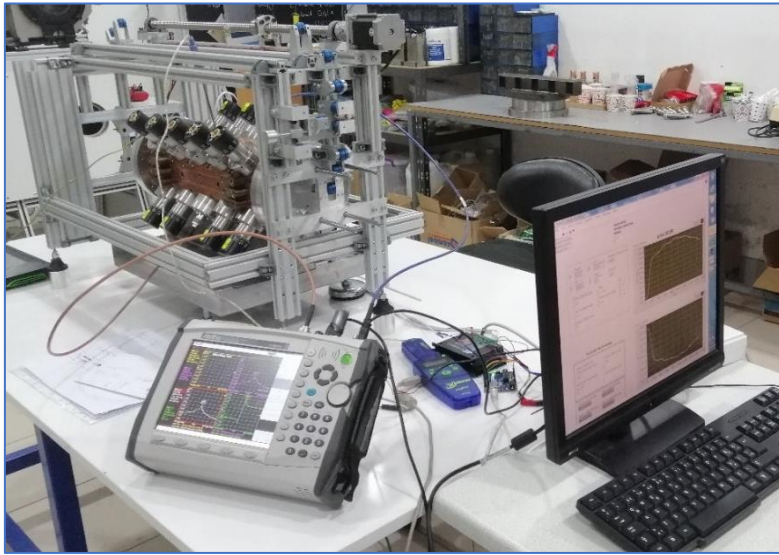
Figure 4: The experimental setup for the bead-pull measurements of the PTAK-RFQ module 0 at KAHVELab.

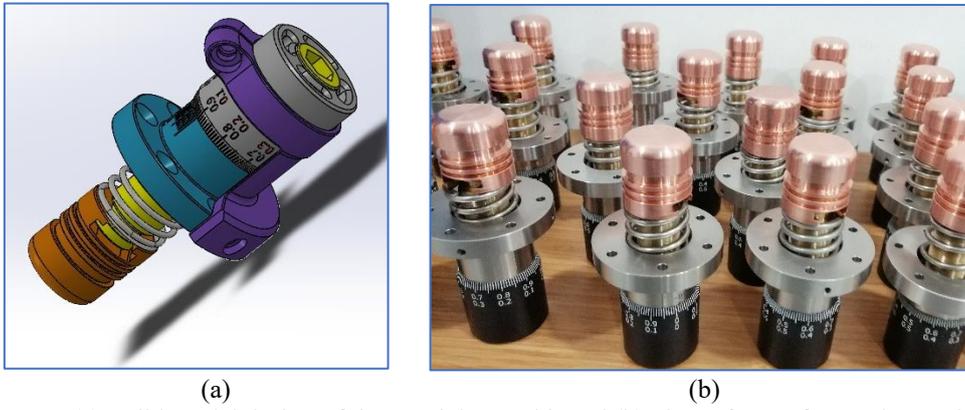
(a) (b)
Figure 5: (a) Solid model design of the special tuner kit and (b) view of manufactured tuner kits.

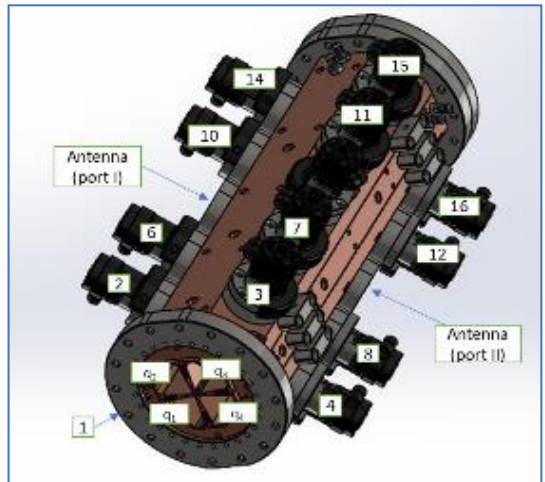
Figure 6: The layout of labeled quadrants and tuners for the PTAK-RFQ module 0.

The raw phase $\phi$ measurements of the transmission coefficient $S_{21}$ measured through bead-pull measurement of each quadrant of the PTAK-RFQ module 0 were performed to get each of the field components. These measurements were aligned in the horizontal and vertical planes using a few data processing steps and then smoothed by the Kernel regression [20-22]. The relative quadrant amplitudes $\beta_1$, $\beta_2$, $\beta_3$, $\beta_4$ were established by taking the square root of the magnetic field in each quadrant which is assigned by proper signs +, -, +, - to take alternating field orientation of $TE_{210}$ mode into account [13].

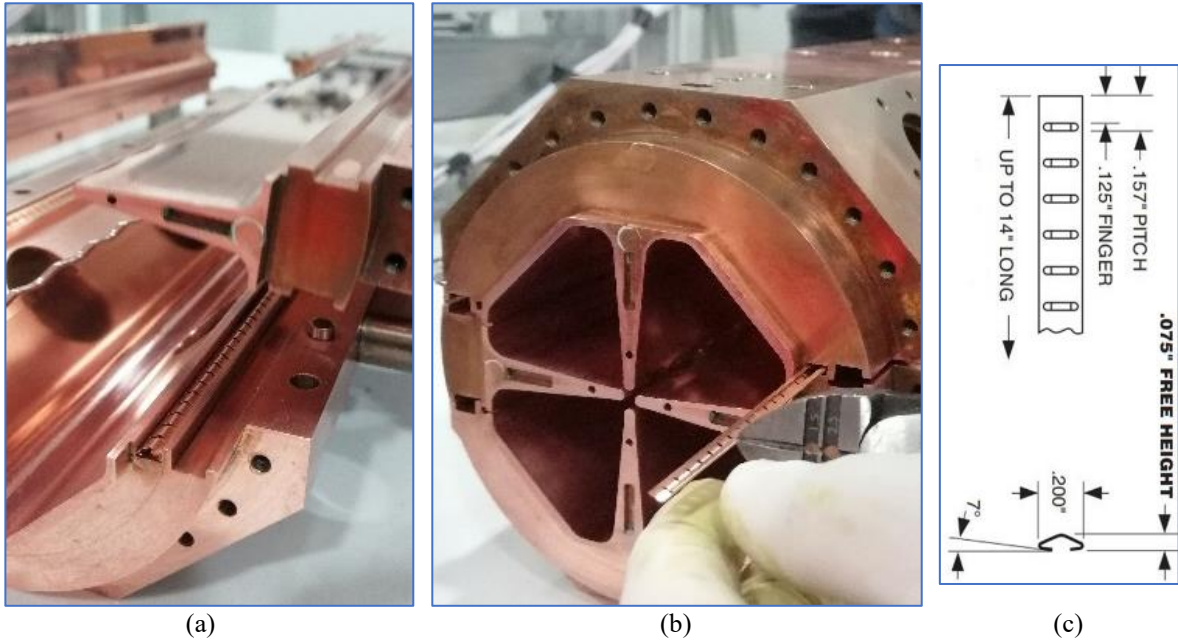

(a) (b) (c)

Figure 7: (a), (b) The Copper C-shaped finger-type RF shields in channels at the junction points of each of the conductive vanes of the PTAK-RFQ module 0, and (c) the dimensions of the RF shield.

The magnetic field distribution at each longitudinal location is defined in terms of the quadrupole field amplitude ($Q^{amp}$) and dipole field amplitudes ($D(S)^{amp}, D(T)^{amp}$). The field amplitude of the quadrupole mode is proportional to

$$Q^{amp} = \frac{\beta_1 - \beta_2 + \beta_3 - \beta_4}{4} \quad (1).$$

Field amplitudes for the dipole modes are proportional to

$$D(S)^{amp} = \frac{\beta_1 - \beta_3}{2} \text{ and } D(T)^{amp} = \frac{\beta_2 - \beta_4}{2} \quad (2).$$

For the PTAK-RFQ module 0 to be properly tuned, it is also necessary to ensure flux conservation $\beta_1 + \beta_2 + \beta_3 + \beta_4 = 0$ during all bead-pull measurements [17].

*2.2. Measurements for the PTAK-RFQ Module 0*

2.2.1. Repeatability of Field Measurement

A series of bead pull measurements were carried out (all tuners were in flush position) to check the repeatability of measurements. It was observed no large deviations between each of the measured $Q^{amp}$, $D(S)^{amp}$, and $D(T)^{amp}$ field amplitudes and the mean of each of the measured field amplitudes as seen in Fig. 8.

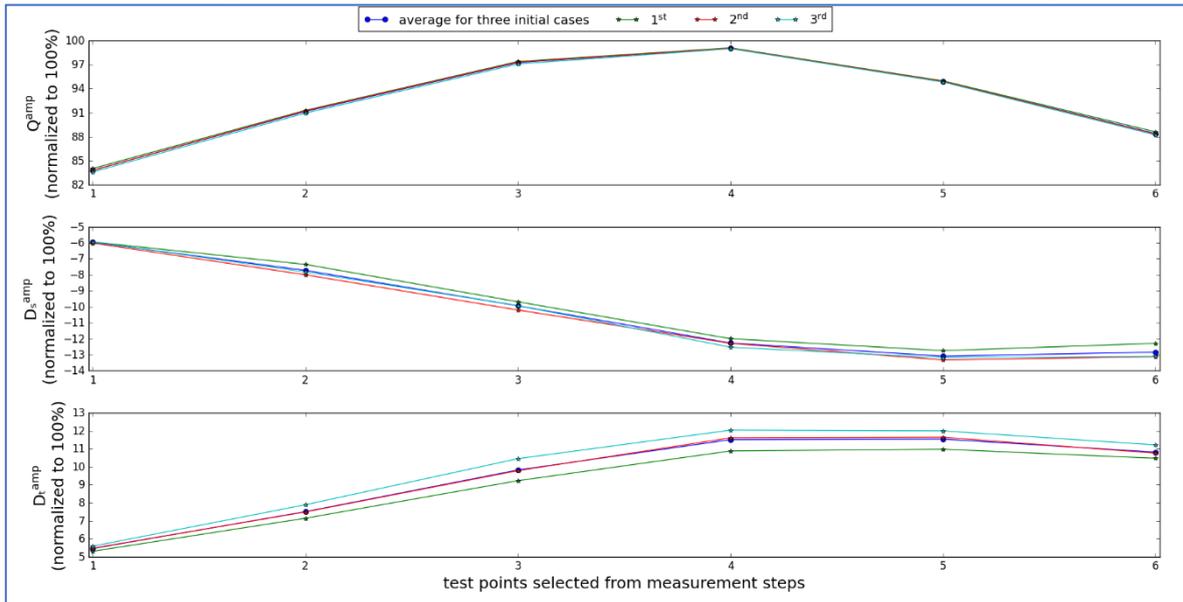

Figure 8: Comparison between each $Q^{amp}$, $D(S)^{amp}$, and $D(T)^{amp}$ field amplitudes obtained from three measurements and the mean of each of the measured field amplitudes for longitudinal field test points.

### 2.2.2. Measured Field Components before the Field Tuning and Simulated Field Components

A comparison between the mean of each of the measured $Q^{amp}$, $D(S)^{amp}$, and $D(T)^{amp}$ field amplitudes before the field tuning and each of the simulated field amplitudes (all tuners were in flush position) is shown in Fig. 9.

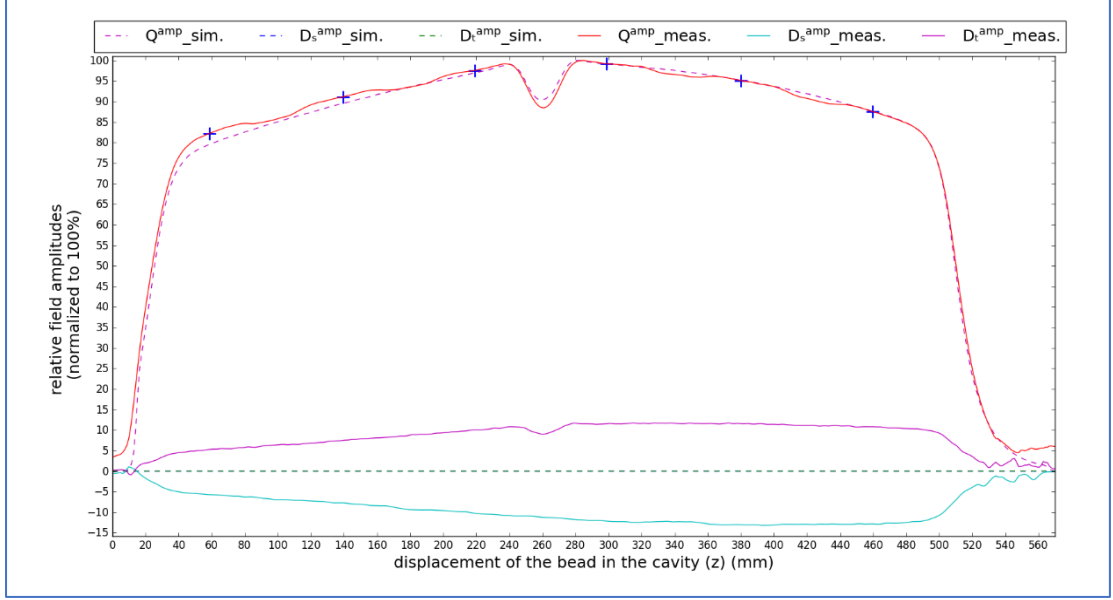

Figure 9: Comparison between the mean of each of the measured $Q^{amp}$, $D(S)^{amp}$, and $D(T)^{amp}$ field amplitudes before the field tuning and each of the simulated field amplitudes.

It can be seen in Fig. 9 that there is a good agreement between the measured quadrupole field amplitude and the simulated quadrupole field amplitude, whereas each of the measured dipole field amplitudes has a remarkable deviation compared to each of the simulated dipole field amplitudes. To achieve the desired field distribution, the measured dipole field amplitudes should be as close to zero as possible, while the smaller deviation in the measured quadrupole field amplitude, compared to the simulated quadrupole field amplitude should be minimized as much as possible, as seen in Fig. 9.

### 2.2.3. Optimization of the Field Tuning Algorithm Developed by CERN

The PTAK-RFQ module 0 was tuned with an optimized version of the tuning algorithm developed by CERN for the HF-RFQ [12,13]. The tuning algorithm for the PTAK-RFQ module 0 is described with a slightly different notation.

The magnetic field distributions at the longitudinal measurement locations, i.e., longitudinal field test points are defined as a vector [12]. The measured actual field distribution $V_{current}$ is defined as the vector of quadrupole field amplitude $Q^{amp}_{current,k}$ and dipole field amplitudes $D(S)^{amp}_{current,k}$ $D(T)^{amp}_{current,k}$ at longitudinal field test points $k = 1,2 ...,$ p. The desired field distribution $V_{target}$ is defined as the vector of quadrupole field amplitude $Q^{amp}_{target,k}$ and dipole field amplitudes $D(S)^{amp}_{target,k}$, $D(T)^{amp}_{target,k}$, $k = 1,2 ...,$ p. The difference between the target and the actual field distribution at the longitudinal field test points k can be easily calculated and it is written as:

$$\Delta V = V_{target} - V_{current} \qquad (3).$$

The $T_{target}$ and $T_{current}$ denote the targeted tuner positions to correct magnetic field deviations and the measured actual tuner adjustments, respectively. It can be defined corrective tuner movements $\vec{\Delta T}$ to be applied to obtain desired field amplitudes by minimizing deviations in each of the current field amplitudes and it is written as:

$$\Delta T = T_{target} - T_{current} \qquad (4).$$

It can be measured both the change of magnetic field distribution $\partial V_{current,k}$ and the change of tuner adjustment $\partial T_{current,l}$ by moving individually each tuner (i.e., other tuners remain in flush position). Thus, it is formed the $\partial V_{current,k} / \partial T_{current,l}$ such that the effect of individual tuner movement on any known longitudinal field test point of each of field components $Q^{AMP}$, $D(S)^{AMP}$, and $D(T)^{AMP}$ could be calculated in first-order approximation. The system of equations can be written as:

$$\Delta V = \frac{\partial V_{current,k}}{\partial T_{current,l}} \Delta T \qquad (5)$$

where the tuner tag number is known as $l = 1, 2 \ldots, r$. It was individually each tuner inserted 3 mm into the 0 module and thus, the $\partial T_{current,l}$ (see Eq. (5)) is denoted as -3 mm. Each entry of the response matrix R is created by derivatives (see in Eq. (5)), which are obtained from bead pull measurements and it can be briefly written as:

$$\Delta V = R \Delta T \quad (6).$$

If the tuning algorithm developed by CERN for the HF-RFQ [12, 13] is optimized for the field tuning of the PTAK-RFQ module 0, it emerges the new matrix equation

$$\begin{bmatrix} 80.2 - Q_1^{AMP} \\ 89.8 - Q_2^{AMP} \\ 97.1 - Q_3^{AMP} \\ 99.1 - Q_4^{AMP} \\ 95.1 - Q_5^{AMP} \\ 87.8 - Q_6^{AMP} \\ 0 - D(S)_1^{AMP} \\ 0 - D(S)_2^{AMP} \\ 0 - D(S)_3^{AMP} \\ 0 - D(S)_4^{AMP} \\ 0 - D(S)_5^{AMP} \\ 0 - D(S)_6^{AMP} \\ 0 - D(T)_1^{AMP} \\ 0 - D(T)_2^{AMP} \\ 0 - D(T)_3^{AMP} \\ 0 - D(T)_4^{AMP} \\ 0 - D(T)_5^{AMP} \\ 0 - D(T)_6^{AMP} \end{bmatrix} = \begin{bmatrix} \frac{\partial Q1}{\partial T1} & \frac{\partial Q1}{\partial T2} & \cdots & \cdots & \cdots & \frac{\partial Q1}{\partial T16} \\ \vdots & \ddots & \ddots & \ddots & \ddots & \vdots \\ \frac{\partial Q6}{\partial T1} & \frac{\partial Q6}{\partial T2} & \ddots & \ddots & \ddots & \frac{\partial Q6}{\partial T16} \\ \frac{\partial Ds1}{\partial T1} & \frac{\partial Ds1}{\partial T2} & \ddots & \ddots & \ddots & \frac{\partial Ds1}{\partial T16} \\ \vdots & \ddots & \ddots & \ddots & \ddots & \vdots \\ \frac{\partial Ds6}{\partial T1} & \frac{\partial Ds6}{\partial T2} & \ddots & \ddots & \ddots & \frac{\partial Ds6}{\partial T16} \\ \frac{\partial Dt1}{\partial T1} & \frac{\partial Dt1}{\partial T2} & \ddots & \ddots & \ddots & \frac{\partial Dt1}{\partial T16} \\ \vdots & \ddots & \ddots & \ddots & \ddots & \vdots \\ \frac{\partial Dt6}{\partial T1} & \frac{\partial Dt6}{\partial T2} & \cdots & \cdots & \cdots & \frac{\partial Dt6}{\partial T16} \end{bmatrix} \begin{bmatrix} T1 - 0 \\ T2 - 0 \\ T3 - 0 \\ T4 - 0 \\ T5 - 0 \\ T6 - 0 \\ T7 - 0 \\ T8 - 0 \\ T9 - 0 \\ T10 - 0 \\ T11 - 0 \\ T12 - 0 \\ T13 - 0 \\ T14 - 0 \\ T15 - 0 \\ T16 - 0 \end{bmatrix} \quad (7).$$

Furthermore, it was decided to modify the entries (see Eq. (7)), where longitudinal field test points of the desired quadrupole field component are located in the first 6 rows of the left column matrix. These corrections on the entries of the matrix were made to ensure compatibility between the measured field components and the simulated field components.

By means of the above-mentioned arrangements, it is aimed to calculate possible correct solutions for the field tuning of the PTAK-RFQ module 0 with the help of the new matrix equation in Eq (7).

### 2.2.3.1. The Inverse of Response Matrix by Singular Value Decomposition (SVD) Method

Eq. (6) must be rearranged by multiplying by the inverse of R to calculate the correct tuner adjustment and it can be written as

$$\Delta T = R^\dagger \Delta V \quad (8)$$

where $R^\dagger$ indicates Moore-Penrose pseudo-inverse of matrix R [23] which is not a square matrix (see Eq. (7)). It was indicated that the matrix R is ill-conditioned and the solutions for unknown tuner adjustment of the HF-RFQ were calculated using a special method based on singular value decomposition, which allows to calculate the inverse of the response matrix R [12, 13]. Response matrix R for the PTAK-RFQ module 0 consists of $K = 3p$ rows, where p corresponds to 6 longitudinal field test points and $L = 16$ columns. The response matrix R can be decomposed by the SVD [23, 24] method as

$$R = U \Sigma V^T \quad (9)$$

where the rectangular matrix $R(K, L)$ is transformed into a rectangular diagonal matrix $\Sigma(\sigma_i)$, orthonormal matrices $U(K, K)$ and $V(L, L)$. The diagonal matrix $\Sigma(\sigma_i)$ as the matrix R, is of the size $(K, L)$. The $U(K, K)$ and the $V(L, L)$ are square matrices. The diagonal entries $\sigma_i$ $(i = 1, 2, \ldots, \mathfrak{g})$ in descending order $\sigma_1 \geqq \sigma_2 \cdots \geqq \sigma_\mathfrak{g} > 0$ are called singular values of the matrix R. The $K > L$ (overdetermined system) is a fundamental requirement of the system to consider when deciding on the number of longitudinal field test points to construct the response matrix R. The Moore-Penrose pseudo-inverse $R^\dagger$ of the matrix R in Eq. (8) can be factorized similarly

$$R^\dagger = V \Sigma^\dagger U^T \quad (10)$$

where $\Sigma^\dagger$ is formed by replacing each positive diagonal entry $\sigma_i$ ($i = 1,2,\ldots,\varrho$) of the diagonal matrix $\Sigma(\sigma_i)$ by its reciprocal and then the matrix $\Sigma^\dagger(\frac{1}{\sigma_i})$ is transposed: if the matrix $\Sigma(\sigma_i)$ is of size (K, L), then the matrix $\Sigma^\dagger(\frac{1}{\sigma_i})$ must be of size (L, K). The diagonal matrix $\Sigma^\dagger$ is arranged as

$$\Sigma^\dagger\left(\frac{1}{\sigma_i}\right) = \begin{bmatrix} \underbrace{\begin{matrix} \frac{1}{\sigma_1} & 0 & \cdots & 0 & 0 & 0 & \cdots & 0 \\ 0 & \frac{1}{\sigma_2} & \ddots & 0 & 0 & \ddots & \ddots & \vdots \\ 0 & \ddots & \frac{1}{\sigma_3} & \ddots & \vdots & 0 & \ddots & \vdots \\ \vdots & \ddots & \ddots & \ddots & 0 & \ddots & \ddots & \vdots \\ 0 & 0 & \cdots & 0 & \frac{1}{\sigma_\varrho} & 0 & \cdots & 0 \\ 0 & \cdots & \cdots & 0 & 0 & 0 & \cdots & 0 \\ \vdots & \ddots & \ddots & \vdots & \vdots & \vdots & \ddots & \vdots \\ 0 & \cdots & \cdots & 0 & 0 & 0 & \cdots & 0 \end{matrix}}_{\varrho\ columns \quad K-\varrho\ columns} \left.\begin{matrix} \\ \\ \\ \\ \\ \\ \\ \\ \end{matrix}\right\}\begin{matrix}\varrho\ rows \\ \\ \\ \\ \\ L-\varrho\ rows \\ \\ \end{matrix} \end{bmatrix} \quad (11).$$

If matrix R has singular values that are so close to zero, it is not possible to accurately compute them. For this case, the matrix is called ill-conditioned, such that the reciprocals $\frac{1}{\sigma_i}$ of these singular values that are so close to zero, lead to numerical errors. In the field tuning of the HF-RFQ [12, 13], it was suggested to circumvent this serious problem by setting the largest diagonal entry of the diagonal matrix $\Sigma^\dagger$ and then the next diagonal entries to zero, in consecutive order.

Note that the matrix $\Sigma^\dagger_{M.P.,0} = \Sigma^\dagger$ is denoted the initial Moore-Penrose (M.P.) inverse, whereas setting the first diagonal element to zero leads to $R^\dagger=0$, such that this solution does not cause any tuner movement. Thus, the matrix $\Sigma^\dagger_{M.P.,(L-1)}$ which has only the first diagonal entry $\frac{1}{\sigma_1}$ denotes the last Moore-Penrose. Possible solutions for tuner adjustment are given (in a slightly different notation than in the references [12, 13, 16]) as

$$\Delta T_{svd,t} = V\Sigma_{M.P.,t}^\dagger U^T \Delta V \quad (t = 0,1,\ldots,(L-1)) \tag{12}$$

where possible solutions for tuner adjustment L-1 are yielded from different diagonal matrices $\Sigma^\dagger$ generated by manipulated singular values. It can be adjusted tuners to new tuner positions by choosing the appropriate prediction to obtain the desired field distribution as

$$T_{target,t} = T_{current} + V\Sigma_{M.P.,t}^\dagger U^T \Delta V \tag{13}.$$

Some solutions, which are outside the physical range of tuner motion can be generated and they must be discarded. Each of the compensated field predictions $V_{svd,t}$ is computed using the original response matrix R as
$$\Delta V_{svd,t} = R \Delta T_{svd,t} \quad (t = 0,1,\ldots,(L-1)) \tag{14}$$
and each of the corrected field prediction $V_{target,t}$ is

$$V_{target,t} = V_{current} + \Delta V_{svd,t} \tag{15}$$

where each of the corrected field predictions guides in checking the remaining solutions. For the current tuning step, the corrected field prediction $V_{target,t}$ where the dipole field amplitudes are as close to zero as possible and the quadrupole field amplitude satisfies the simulation results as much as possible should be chosen.

After the tuner adjustment $\Delta T_{svd,t}$, whose is corresponding to the chosen $V_{target,t}$ is applied on the module, a new measurement is performed and it emerges a new vector $V_{current}$. The above-mentioned process is iterated until the measured field components satisfy in terms of their requirements for the desired field distributions with the desired accuracy [12, 13, 16, 17].

Based on the fact that each iteration for field tuning steps is time-consuming, it is satisfactory to use only the initial response matrix R and to choose a solution corresponding to a different value of *t* for the tuner adjustment [12, 13]. It is clear that this method has a remarkable measurement advantage [16].

## 2.3. Tuning Steps for the PTAK-RFQ Module 0

For the first tuning step of the PTAK-RFQ module 0, it is given 16 possible solutions for corrective tuner movements $\Delta T_{svd,t}$ (see in Eq. (12)) and 16 predictions $\Delta V_{svd,t}$ (see in Eq. (14)) that would compensate each of $Q^{amp}$, $D(S)^{amp}$ and $D(T)^{amp}$ field amplitudes after applying each of solutions for the tuner adjustment, shown in Fig. 10.

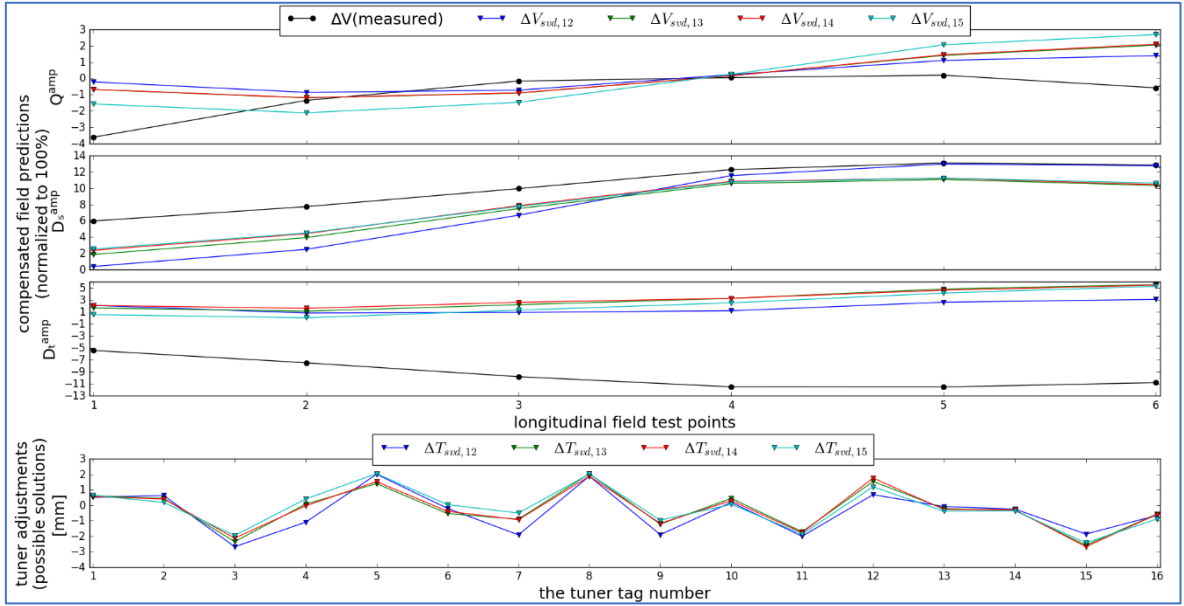

Figure 10: Some of the possible solutions for corrective tuner adjustments $\Delta T_{svd,t}$ and compensated field prediction $\Delta V_{svd,t}$ after applying each of the solutions for the tuner adjustment.

First, the initially measured field components and then the measured field components after the next five tuning steps are shown in Fig. 11. It is also shown the measured field components for 1st frequency tuning in Fig.11.

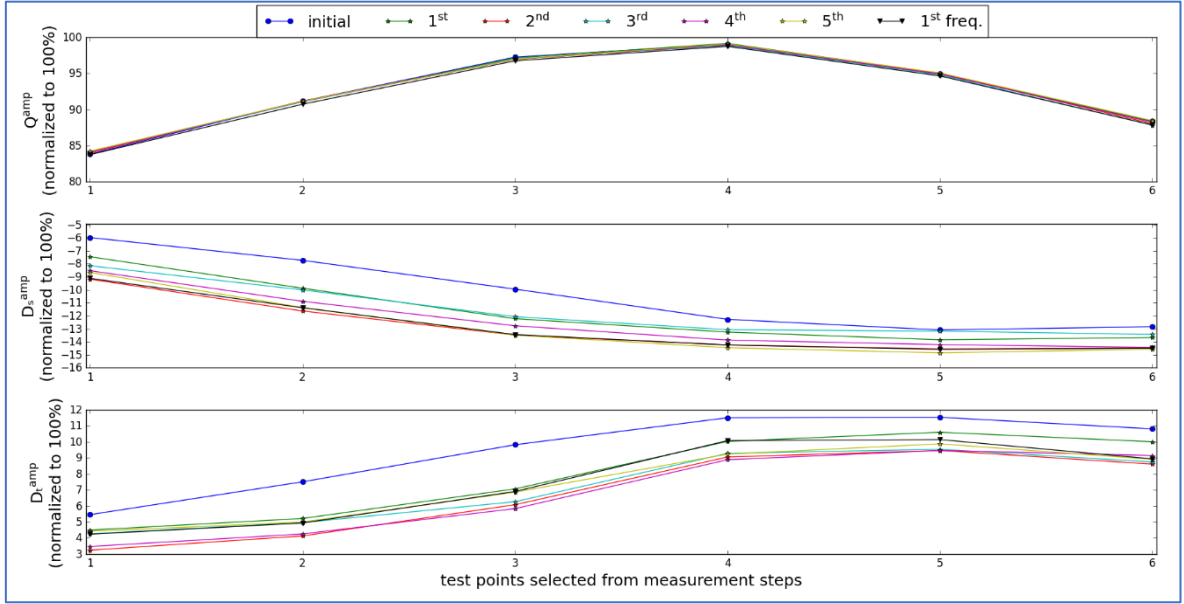

Figure 11: The initially measured field amplitudes, the measured field amplitudes for the next five tuning steps, and then 1st frequency tuning.

The measured $Q^{amp}$ field component had an almost good agreement in the initial case and this agreement remained almost the same for the next 5 tuning steps. Furthermore, the deviation in the measured $D(T)^{amp}$ field component in the initial case, was limited to a very small improvement after the next 5 tuning steps, whereas the deviation in the measured $D(S)^{amp}$ field component in the initial case, increased slightly after the next 5 tuning steps as shown in Fig. 11.

The field tuning process was iterated using the prediction $\Delta T_{svd,15}$ which corresponds to the prediction $\Delta V_{svd,15}$ for the first 8 tuning steps, including the 8th tuning step, and then using the prediction $\Delta T_{svd,14}$ which corresponds to the prediction $\Delta V_{svd,14}$ for 9th tuning step. The measured field components for each of tuning steps 6th-9th and then for 2nd frequency tuning are reported in Fig. 12.

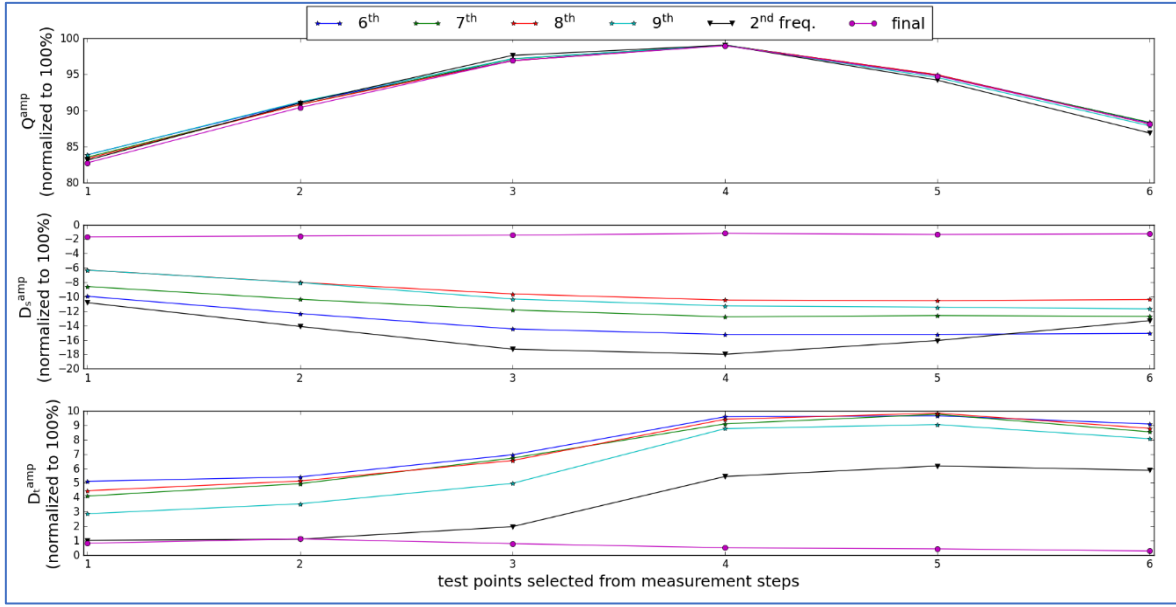

Figure 12: The measured field amplitudes for each of tuning steps 6th to 9th, for the 2nd frequency tuning, and then the measured field amplitudes for the final tuning step.

The deviation in the first two longitudinal field test points in terms of the measured $D(T)^{amp}$ field component was minimized after 2nd frequency tuning step as seen at the bottom of Figure 12. It is known that the corrective action on these longitudinal field test points is around the tuners, which are arranged in the same cross-section (i.e., tuners 1 to 4) as seen in Fig. 6. To minimize the deviation of the measured $D(S)^{amp}$ field component in the final tuning step, the lengths of tuners 2 and 4 were adjusted for tuners 1 and 3 which are arranged at the same horizontal cross-section, respectively. Furthermore, lengths of tuners that are arranged at the same horizontal cross-section (i.e., tuners 1 to 4) of the PTAK-RFQ module 0 were adjusted for tuners that are arranged at other horizontal cross-sections (i.e., tuners 5 to 8 and so on) as shown in Fig. 6, respectively.

After the manual tuning step, a deviation of -600 MHz in the measured frequency was observed. To eliminate this frequency error, all tuners were inserted 2.4 mm into the cavity and then a small deviation of +10 MHz in the measured frequency was observed.

Tuning of the operating frequency and tuning of each of the field components of the PTAK-RFQ module 0 was performed in several iterations. Errors in the initial measured field components, in measured field components after each tuning step, and the corresponding errors in terms of operating frequency are summarized in Table 2. Errors for $Q^{amp}$, $D(S)^{amp}$, and $D(T)^{amp}$ were suppressed to an error of 0.8%, 1.6%, and 1.1%, respectively. It was calibrated to the VNA with a narrower calibration range (798-802 MHz) in the next tuning step and the frequency error was +96 kHz. The operating frequency of the PTAK-RFQ module 0 was finally set to 800.000 MHz by extracting all the tuners by 0.04 mm. The final bead pull measurement was performed with the final tuner adjustment as shown in Figure 13.

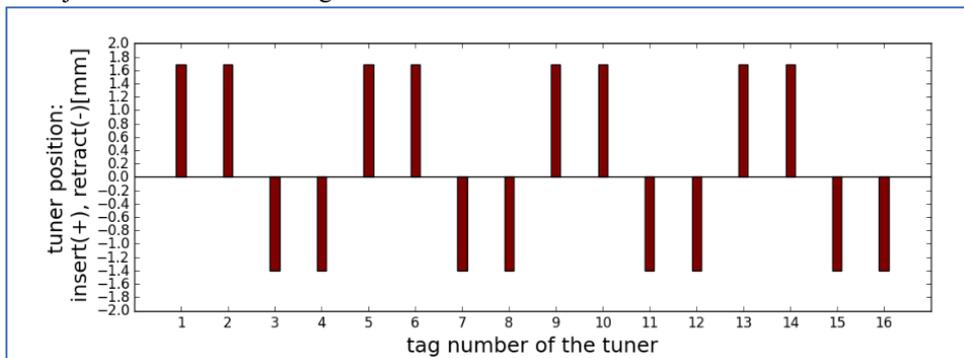

Figure 13: Final tuner positions on the PTAK-RFQ module 0.

Table 2: Errors in the initially measured field components, in the measured field components after each tuning step, and the corresponding errors in terms of operating frequency in chronological order.

| Tuning Steps (Prediction number or Comments) | Deviation from 800 MHz Operating Frequency (kHz) | Max. Deviation from the Desired Field Amplitudes (normalized to 100%) | | |
|---|---|---|---|---|
| | $f$ | $Q^{amp}$ | $D(S)^{amp}$ | $D(T)^{amp}$ |
| Initial | −350 | ±1.6% | ±12.8% | ±11.6% |
| 1st iteration ($T_{svd,15}$) | −10 | ±1.5% | ±13.7% | ±10.6% |
| 2nd iteration ($T_{svd,15}$) | −40 | ±1.6% | ±14.5% | ±9.5% |
| 3rd iteration ($T_{svd,15}$) | −90 | ±1.5% | ±13.4% | ±9.5% |
| 4th iteration ($T_{svd,15}$) | −110 | ±1.5% | ±14.4% | ±9.5% |
| 5th iteration ($T_{svd,15}$) | −200 | ±1.5% | ±14.6% | ±9.9% |
| 1st freq. tuning (All tuners inserted (+) 2 mm to the cavity) | +240 | ±1.1% | ±14.5% | ±10.1% |
| 6th iteration ($T_{svd,15}$) | +210 | ±1.5% | ±15.2% | ±9.7% |
| 7th iteration ($T_{svd,15}$) | +220 | ±1.2% | ±12.8% | ±9.8% |
| 8th iteration ($T_{svd,15}$) | +170 | ±1.2% | ±10.5% | ±9.8% |
| 9th iteration ($T_{svd,14}$) | +320 | ±1.6% | ±11.2% | ±9% |
| 2nd freq. tuning (All tuners inserted (+) 0.15 mm to the cavity) | −160 | ±1.4% | ±18% | ±6.2% |
| Adjustment without any prediction | −620 | No meas. | No meas. | No meas. |
| 3rd freq. tuning (All tuners inserted (+) 2.40 mm to the cavity) | +10 | ±0.8% | ±1.6% | ±1.1% |
| A new calibration of VNA (range: 798-802 MHz) | +96 | No meas. | No meas. | No meas. |
| Final adjustment (All tuners extracted (-) 0.04 mm from the cavity) | 0 | ±0.8% | ±1.6% | ±1.1% |

Fig. 14 shows the comparison between simulated $Q^{amp}$, $D(S)^{amp}$, and $D(T)^{amp}$ field components and measured $Q^{amp}$, $D(S)^{amp}$, and $D(T)^{amp}$ field components after the final tuning step.

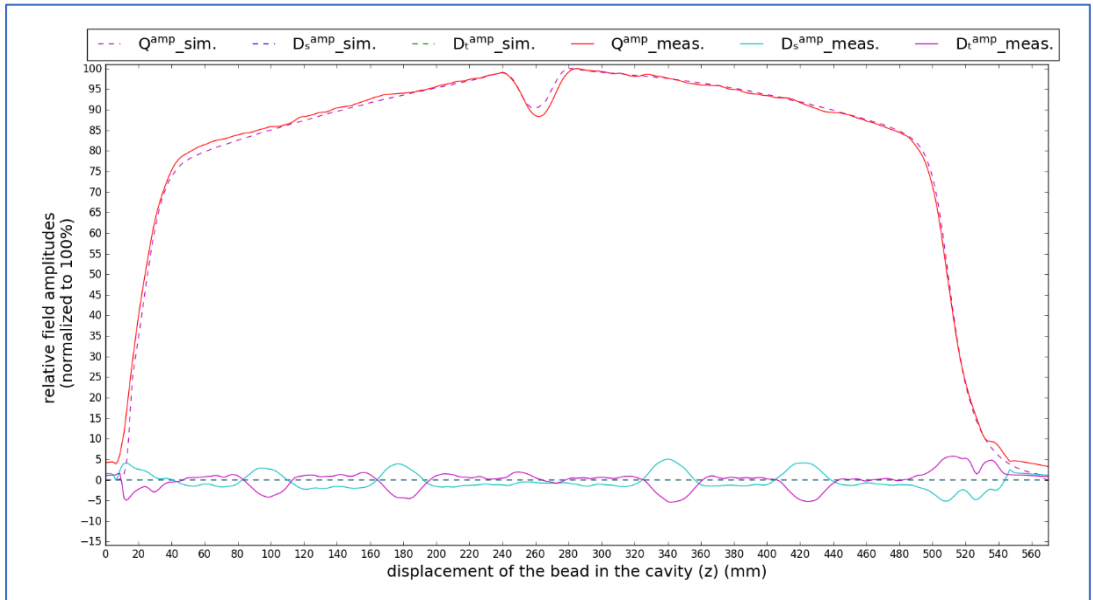

Figure 14: The Comparison between simulated field components and measured field components after the final tuning.

The errors in the frequency and the field after the final tuning step fulfill adequately the desired requirements for the field and the frequency tuning of the PTAK-RFQ module 0.

The thermal behavior of the PTAK-RFQ module 0 from the ordinary copper material is dominated by the ambient temperature, which changes rapidly during the day. Some errors, which originate from the response matrix based on the measurements, occur as the phase shift due to the effect of temperature changing over time. To avoid such measurement errors, the ambient temperature must be kept as constant as possible throughout the bead-pull measurement process. In addition to the ambient temperature, it is known that air humidity also causes errors in terms of frequency [13, 16].

The ambient temperature and the air humidity data were recorded during all bead-pull measurements of the PTAK-RFQ module 0. Note that despite the known determinant influences of the ambient temperature and the air humidity on the final field distribution and the operating frequency of the PTAK-RFQ module 0, these influences were not taken into account in all tuning steps of the PTAK-RFQ module 0.

## 3. The Initial Measurements for Quality Factors of the PTAK-RFQ Module 0

The four-vane PTAK-RFQ module 0 was assembled by using finger-type RF shields, which are placed in channels between each conductive vane to resist RF leakages. Therefore, the initial measurements for the quality factors have been carried out to observe the effect of finger-type RF shields on the quality factors. Ends of the PTAK-RFQ module 0 were terminated by an aluminum cap and aluminum extension tube to obtain simulated boundary conditions. The initial quality factors were measured despite the non-delivery of finger-type RF shields to be placed in channels on ends of the PTAK-RFQ module 0 (Fig. 2(c)).

The VNA (50Ω) was connected to the input port and the output port on the PTAK-RFQ module 0. The transmission-type measurements were performed to determine the unloaded quality factor $Q_0$ of the PTAK-RFQ module 0. The weak coupling ($|S_{21}| \approx -40\ dB$ at resonance) was used due to the coupling losses which are caused by two pick-up antennas (50Ω) of unequal size, such that unloaded quality factor ($Q_0$) $\approx$ loaded quality factor ($Q_l$) [25]. The loaded Q-factor $Q_l$ is calculated by measuring frequencies at drop points of 3 dB bandwidth of the transmission coefficient $S_{21}$ and the center frequency $f_0$. The loaded quality factor $Q_l$ is calculated as

$$Q_l = \frac{f_0}{f^+_{-3dB} - f^-_{-3dB}} \quad (16)$$

[26].

The quality factor measurements were carried out while all tuners were in a flush position on the PTAK-RFQ module 0. The measured unloaded quality factor of the PTAK-RFQ module 0 was found to be approximately 50% of the simulated unloaded quality factor without using the RF fingers.

## CONCLUSION

The final 800 MHz four-vane PTAK-RFQ has been built recently from OFE-Cu (Oxygen-free electric copper) material at the local manufacturing facilities in Turkey. All the developments and solutions that emerged during the mechanical, vacuum, RF tests, and fabrication processes of the PTAK-RFQ module 0 will also facilitate the commissioning of the final PTAK-RFQ.

The tuning algorithm developed by CERN has been optimized by the PTAK-RFQ module 0. This tuning algorithm will be augmented by including new entries to the system of equations for extra test points and extra tuners of the final PTAK-RFQ which consists of two modules.

A minor design flaw (see bottom right of Fig 14) in terms of the field distribution was reported as a result of the end cap design to obtain boundary conditions in simulation and it is negligible for the PTAK-RFQ module 0. This minor design flaw was revised for the design of the final PTAK-RFQ.

It has been observed that finger-type RF shields have substantial resistance to RF leaks in the PTAK-RFQ module 0 during quality factor measurements. Measured quality factors are expected to be consistent with simulated quality factors after the assembly of the undelivered finger-type RF shields.

The ambient temperature and air humidity data will be recorded during bead pull measurements of the final PTAK-RFQ and these data will be taken into account in all field and frequency tuning steps.

## ACKNOWLEDGEMENTS

We thank TENMAK (Turkish Energy Nuclear and Mineral Research Agency) for providing the He Leak test device. We wish to thank Birant Baran and DORA MAKINA CORP. for their valuable support and contributions to achievement of the desired operational settings of the PTAK-RFQ module 0.

This project is supported by The Scientific and Technological Research Council of Turkey (TUBITAK) Project no: 118E838. It was also provided equipment support by Boğaziçi University Scientific Research Project no: 16B03S10.